\documentclass[aip,jcp,reprint,groupedaddress,twocolumn]{revtex4}

\usepackage{amsmath}    
\usepackage{amssymb}
\usepackage{graphicx}   
\usepackage{graphics}
\usepackage{hyperref}   
\usepackage{comment}
\usepackage{color}
\usepackage{epstopdf}
\usepackage[caption=false]{subfig}

\def\bx{{\mathbf{x}}}

\newcommand{\Ektil}{\widetilde{E}_k}

\begin{document}

\title{Discrete discontinuous basis projection method for large-scale electronic structure calculations}
\author{Qimen Xu and Phanish Suryanarayana}
\email[Email: ]{phanish.suryanarayana@ce.gatech.edu}
\affiliation{College of Engineering, Georgia Institute of Technology, Atlanta, GA 30332, USA}
\author{John E. Pask}
\affiliation{Physics Division, Lawrence Livermore National Laboratory, Livermore, CA 94550, USA}
\date{\today}

\begin{abstract}
We present an approach to accelerate real-space electronic structure methods several fold, without loss of accuracy, by reducing the dimension of the discrete eigenproblem that must be solved. To accomplish this, we construct an efficient, systematically improvable, discontinuous basis spanning the occupied subspace and project the real-space Hamiltonian onto the span. In calculations on a range of systems, we find that accurate energies and forces are obtained with 8--25 basis functions per atom, reducing the dimension of the associated real-space eigenproblems by 1--3 orders of magnitude.
\end{abstract}


\maketitle

Over the course of the past few decades, quantum mechanical calculations have become a linchpin of materials research by virtue of the fundamental insights they provide and predictive power they afford. Of the many formulations employed, Kohn-Sham density functional theory (KS-DFT) \cite{Kohn1965} has become the most widely used in practice due to its generality, relative simplicity, and high accuracy-to-cost ratio \cite{burke2012dft,becke2014dft}. However, while less costly than wavefunction based alternatives, the solution of the Kohn-Sham equations remains a formidable task, severely restricting the range of physical systems that can be investigated. These restrictions become even more acute in molecular dynamics investigations \cite{burke2012dft} wherein the Kohn-Sham equations may be solved tens or hundreds of thousands of times to reach the time scales relevant to phenomena of interest.

The planewave pseudopotential method \cite{Martin2004} has been among the most widely used methods for the solution of the Kohn-Sham equations. By virtue of its Fourier basis, the method is accurate, simple to use since it relies on a single convergence parameter, and highly efficient on moderate computational resources employing well optimized Fast Fourier Transforms (FFTs). However, the use of a Fourier basis limits the method to periodic boundary conditions so that finite systems, such as molecules and clusters, as well as semi-infinite systems, such as nanotubes and surfaces, require the introduction of artificial periodicity with large vacuum regions between periodic replicas. This limitation also necessitates a neutralizing background density to avoid Coulomb divergences when treating charged systems. Moreover, the method's reliance on Fourier transforms hampers scalability on parallel computing platforms, thus limiting the system sizes and time scales that can be reached.

To overcome the limitations of the planewave method, a number of approaches employing systematically improvable, localized representations have been developed over the last two decades \cite{arias1999wav,beck2000rsmeth,pask2005femeth,saad2010esmeth,lin2012adaptive}. Of these, perhaps the most mature and widely used to date are the finite-difference methods \cite{beck2000rsmeth,saad2010esmeth}. These methods maximize computational locality by discretizing all quantities of interest on a real-space grid. In so doing, both Dirichlet and periodic boundary conditions can be accommodated, thus enabling efficient and accurate treatment of finite, semi-infinite, and charged systems, as well as bulk 3D systems. Moreover, convergence is controlled by a single parameter and large-scale parallel computational resources can be efficiently leveraged by virtue of the method's simplicity, locality, and freedom from communication-intensive transforms such as FFTs. However, the large number of grid points required to obtain accurate energies and forces ($\sim$400--30,000 per atom), and lack of efficient preconditioners for iterative solutions have hampered the competitiveness of real-space methods in practice.

Significant advances in recent years have accelerated real-space methods substantially. Since pioneering work in the 1990s \cite{bernolc1991mg,chel1994fdpp,briggs1995mg,seits1995fdcg,gygi1995real}, the number of grid points required to obtain accurate energies and forces has been notably reduced by double-grid techniques \cite{OnoHir99}, ultrasoft pseudopotential \cite{hodak2007rsuspp} and projector augmented wave \cite{mort2005rspaw} formulations, high-order integration \cite{BobSchChe15}, and nonlocal force reformulation \cite{ghosh2016sparc1,ghosh2016sparc2}. Moreover, the lack of efficient preconditioners has been resolved by replacing traditional iterative solvers with Chebyshev-polynomial filtered subspace iteration (CheFSI) \cite{zhou2006self}, eliminating the need for preconditioning entirely. With these and other advances, real-space methods have been applied to systems containing thousands of atoms \cite{alamany2008fdlarge}, and have now outperformed established planewave codes in applications to both finite \cite{ghosh2016sparc1} and extended \cite{ghosh2016sparc2} systems.

In this work, we present an approach to accelerate current state-of-the-art real-space electronic structure methods several fold, without loss of accuracy, by systematically reducing the cost of the key computational step: the determination of Kohn-Sham orbitals spanning the occupied subspace.\footnote{Since our aim is a general method, applicable to metals and insulators alike, we consider the standard $\mathcal{O}(N^3)$ formulation, though the approach here will be advantageous in the $\mathcal{O}(N)$ context as well.}  We do this by systematically and substantially reducing the dimension of the discrete eigenproblem that must be solved. To accomplish this, we leverage a recent breakthrough in electronic structure methodologies: discontinuous basis sets \cite{lin2012adaptive}. By allowing the basis to be discontinuous, it can be simultaneously strictly local, orthonormal, systematically improvable, and highly efficient, requiring just tens of basis functions per atom to attain chemical accuracy (mHa/atom and mHa/Bohr in energy and forces, respectively). We construct such a basis spanning the occupied subspace to desired accuracy by solving local Kohn-Sham problems on the same grid as used for the global Kohn-Sham problem. The result is a discrete, discontinuous basis spanning the occupied subspace of the real-space Hamiltonian. We then project the large, sparse Hamiltonian onto the subspace, thus reducing its dimension substantially without loss of accuracy in the relevant subspace. Upon solving the reduced problem, the full-matrix eigenvalues and vectors are readily recovered and the remainder of the real-space calculation proceeds unmodified. The method is thus straightforward to implement and use. Having these key ingredients, we refer the approach as discrete discontinuous basis projection (DDBP). In calculations of quasi-1D, quasi-2D, and bulk metallic systems, we find that accurate energies and forces are obtained with 8--25 projection basis functions per atom, reducing the dimension of full-matrix eigenproblems by 1--3 orders of magnitude. 


We consider the solution of the Kohn-Sham equations for a system of $N_a$ atoms in a domain $\Omega$ discretized on a uniform grid $\mathcal{K} = \bigcup_{q = 1}^{N_d} \{ \bx_q \}$ in real space. In each iteration of the self-consistent field (SCF) solution, the following linear eigenproblem must be solved:
\begin{equation} \label{Eq:Eigenproblem:FD}
\mathcal{H}_d \psi_n = \lambda_n \psi_n \,, \quad n=1,2, \ldots, N_{s} \,,
\end{equation}
where $\mathcal{H}_d$ is the discrete Hamiltonian, $\psi_n$ are the discrete Kohn-Sham orbitals with energies $\lambda_n$, and $N_s$ is the number of occupied states. This constitutes a large, sparse eigenproblem of size $N_d \times N_d$, where $N_d \gg N_s$. As the system size grows, the solution of this problem becomes the bottleneck, regardless of discretization or solution strategy employed. 

To reduce the dimension of the real-space eigenproblem \eqref{Eq:Eigenproblem:FD}, we begin by constructing an efficient, systematically improvable basis spanning the occupied subspace to desired accuracy. 
In order to obtain strict locality and orthonormality, we compute an adaptive local basis \cite{lin2012adaptive} in real space. For this purpose, the domain $\Omega$ is partitioned into non-overlapping elements: ${\Omega = \bigcup_{k = 1}^{N_E} E_k}$, with the collection of grid points encompassed by $E_k$ denoted by $\mathcal{K}_k$. In doing so, ${\mathcal{K} = \bigcup_{k = 1}^{N_E} \mathcal{K}_k}$ with $\mathcal{K}_k \cap \mathcal{K}_{k'} = \emptyset$ for $k \neq k'$. The extended element $\Ektil$ is defined to be the uniform extension of $E_k$ in all directions, with $\Ektil \setminus E_k$ referred to as the \textit{buffer region}. \footnote{If opposite faces of $E_k$ coincide with the boundary of $\Omega$, no buffer is employed in that direction and the boundary conditions on $\Ektil$ in that direction are the same as those on $\Omega$.}


\begin{figure}
\includegraphics[keepaspectratio=true,width=0.485\textwidth]{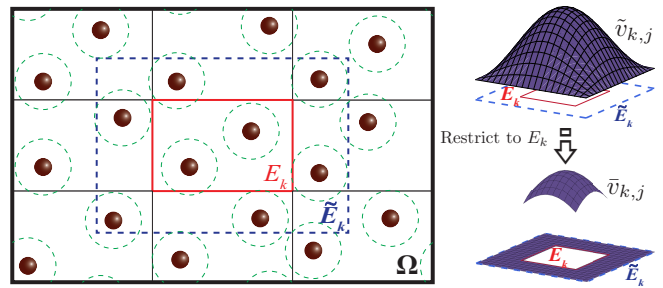} 
\caption{\label{Fig:DDP:Schematics} Discrete discontinuous basis construction. Kohn-Sham eigenfunctions $\widetilde{v}_{k,j}$ are computed in real space in extended elements $\Ektil$, restricted to elements $E_k$ to form strictly local $\bar{v}_{k,j}$, and orthonormalized to form strictly local, orthonormal, discrete discontinuous basis $v_{k,j}$.}
\end{figure}

As illustrated in Fig.~\ref{Fig:DDP:Schematics}, the construction of the basis during an SCF iteration proceeds as follows. 
First, we solve the Kohn-Sham equations in each extended element $\Ektil$, yielding discrete equations
\begin{equation} \label{Eq:BasisSolution}
\widetilde{\mathcal{H}}_k \widetilde{v}_{k,j} = \widetilde{\lambda}_{k,j} \widetilde{v}_{k,j} \,, \quad j=1,2, \ldots, N_{b,k} \,,
\end{equation}
where $N_{b,k}$ is the number of basis functions in $\Ektil$ and $\widetilde{\mathcal{H}}_k$ is formed by restricting the potential in $\Omega$ to $\Ektil$, with either periodic or Dirichlet boundary conditions on the Laplacian. Second, we restrict the basis so obtained on $\Ektil$ to $E_k$ by truncating it outside $E_k$, i.e.,
\begin{equation} \label{Eq:BasisFuncRestriction}
 \bar{v}_{k,j}(\bx_q) = \left\{
 					\begin{array}{ll}
 						\widetilde{v}_{k,j}(\bx_q) & \quad \bx_q \in \mathcal{K}_k \,, \\
 						0 & \quad \text{otherwise} \,.
 					\end{array}
 				\right.
\end{equation}
Third, we orthonormalize the basis in $E_k$: 
\begin{equation} \label{Eq:Orth}
v_{k,j} = \sum_{j=1}^{N_{b,k}} c_{k,j} \bar{v}_{k,j} \,,
\end{equation} 
where $c_{k,j}$ are chosen such that $v_{k,j}$ are orthonormal within $E_k$. Since $ v_{k,j}$ has zero overlap with $ v_{k',j'}$ for $k \neq k'$, the basis $v_{k,j}$, $k=1,2,\ldots,N_E$, $j=1,2,\ldots,N_{b,k}$, is orthonormal on the global domain $\Omega$. 

Next, we project the Hamiltonian $\mathcal{H}_d$ onto the subspace spanned by the discrete, strictly local, orthonormal basis generated above to arrive at the following reduced eigenproblem:
\begin{equation}\label{Eq:ProjectedEigenproblem}
\mathcal{H}_p w_n = \mu_n w_n \,, \quad n=1,2, \ldots, N_{s} \,,
\end{equation}
where $\mathcal{H}_p = V^T \mathcal{H}_d V$, with $V$ denoting the matrix whose columns are the basis vectors $v_{k,j}$, $k=1,2,\ldots,N_E$, $j=1,2,\ldots,N_{b,k}$. The dimension of this reduced problem is $N_b \times N_b$ with $N_b = \sum_{k=1}^{N_E}  N_{b,k}$, which is significantly smaller than the dimension of the original real-space eigenproblem \eqref{Eq:Eigenproblem:FD}. 

Finally, upon solution of the reduced problem \eqref{Eq:ProjectedEigenproblem}, the corresponding full-matrix vectors are readily recovered as ${\psi_n = V w_n}$, whereupon the remainder of the real-space calculation proceeds unmodified.

Obtaining the reduced problem from the discrete real-space Hamiltonian rather than from the differential operator yields significant advantages. First, because continuity and boundary conditions are already incorporated in the discrete Hamiltonian, the discrete basis need not satisfy either: it must merely span the desired subspace. In contrast, when discretizing the differential operator in a discontinuous basis, continuity and boundary conditions are incorporated via penalty and surface terms, thus requiring surface integrals to be computed and penalty parameters to be set controlling the degree of continuity at interelement boundaries \cite{lin2012adaptive}. Second, the reduced Hamiltonian obtained by the present approach has significantly smaller spectral width than the full Hamiltonian, thus accelerating iterative solution methods. In contrast, when discretizing the differential operator directly, the penalty parameter must be sufficiently large to ensure the stability of the approximation \cite{lin2012adaptive}, leading to larger spectral widths than for corresponding planewave and finite-difference discretizations \cite{banerjee2016chebyshev}. Finally, in the present approach, all computations are carried out on a single, uniform real-space grid; whereas when discretizing the differential operator, separate discretization and quadrature grids are employed, making convergence less straightforward.

We now consider the reduction in computational cost afforded by DDBP. For this purpose, let us denote the finite-difference order by $n_o$ and the partition dimension by $D$.  In addition, let $\hat{N}_d = N_d/N_a$, $\hat{N}_b = N_b/N_a$, and $\tilde{N}_a = N_a/N_E$. The asymptotic cost of the electron density calculation using iterative diagonalization \cite{davidson1975iterative,zhou2006self} is a factor $(\hat{N}_d/\hat{N}_b)$ smaller for the reduced eigenproblem. Whereas for systems of moderate size, where matrix-vector multiplies dominate computational cost, the efficiency is determined by the number of nonzeros in the Hamiltonian, which is a factor $(3n_o+1) \hat{N}_d /(2D+1) \hat{N}_b^2 \tilde{N}_a$ smaller for the reduced problem.\footnote{The nonlocal projectors are typically applied in matrix-free fashion, with the associated cost being a factor of $\mathcal{O}\left(r_c \hat{N}_d/\hat{n}_{pe}\hat{N}_b \tilde{N}_a\right)$ smaller for the reduced problem, where $r_c$ is the ratio of the average number of grid points within the support of each projector to $\hat{N}_d$ and $\hat{n}_{pe}$ is the average number of elements spanned by nonlocal projectors.} This translates to speedups of $\mathcal{O}\left((3n_o+1) \hat{N}_d^2 /(2D+1) \hat{N}_b^3 \tilde{N}_a \right)$ for linear-scaling methods \cite{suryanarayana2017sqdft} in which the density matrix is expressed directly in real space.
 
As we show below, both $\tilde{N}_a$ and $\hat{N}_b$ are $\mathcal{O}(10)$ in practice. Therefore, the prefactor associated with DFT calculations for both diagonalization and linear-scaling methods is significantly reduced. This is a consequence of (i) the reduced number of operations when starting from $\mathcal{H}_p$ compared to $\mathcal{H}_d$, as discussed above, (ii) the smaller spectral width of $\mathcal{H}_p$ compared to $\mathcal{H}_d$, (iii) the block nature of $\mathcal{H}_p$, which enables efficient dense linear algebra, and (iv) the minimal wall time for basis generation, which scales linearly with system size and is naturally parallel. Overall, the above estimates suggest that DDBP may provide up to an order of magnitude speedup for diagonalization based methods and up to two orders of magnitude speedup for linear-scaling methods.


To demonstrate the accuracy and efficiency of DDBP, we implemented the approach in the M-SPARC prototype code, a serial matlab implementation of the large-scale parallel real-space code, SPARC \cite{ghosh2016sparc1,ghosh2016sparc2}. In order to demonstrate the generality of the approach, and particular efficiency in lower-dimensional calculations, we consider three systems with distinct electronic structure and dimensionality: a (3,0) carbon nanotube (CNT) with bond length $2.68$ Bohr, a buckled silicene sheet with bond length $4.25$ Bohr, and fcc aluminum with lattice constant $7.78$ Bohr. The atoms are perturbed by up to ten percent of the bond length, as typical in relaxation and dynamics. In all  calculations, we employ norm-conserving Troullier-Martins pseudopotentials \cite{Troullier}, the local density approximation (LDA) with Perdew-Wang parametrization \cite{perdew1992accurate}, and Fermi-Dirac smearing of $10^{-3}$ Ha to accommodate partial occupation.

In the real-space method, we employ a twelfth order central finite-difference approximation with mesh-sizes of $0.20$, $0.40$, and $0.65$ Bohr for the CNT, silicene, and aluminum systems, respectively. The resulting discretization errors in the energy and atomic forces are within mHa/atom and mHa/Bohr, respectively, as typical in applications. 
In the DDBP method, we perform a $D=1, 2, \text{ and } 3$ dimensional partition with $\tilde{N}_a \sim 12, 4, \text{ and } 4$ atoms/element for the CNT, silicene, and aluminum systems, respectively. Equations~\eqref{Eq:Eigenproblem:FD}, \eqref{Eq:BasisSolution}, and \eqref{Eq:ProjectedEigenproblem} are subject to periodic boundary conditions and solved via Chebyshev-polynomial filtered subspace iteration \cite{zhou2006self}, with orthonormalization via Cholesky factorization \cite{banerjee2016chebyshev}. The SCF iteration is accelerated using the Periodic Pulay method\cite{banerjee2016periodic}. 

In the construction of the DDBP basis, the buffer region must be sufficiently large to minimize boundary condition effects within elements but not so large as to introduce contributions from distant atoms into the local basis. In practice, an efficient buffer size can be determined for a given system by a series of calculations on representative smaller systems. In so doing, one generally finds that the accuracy of the basis is insensitive to the specific choice beyond $\sim$5 au. Since for our prototype serial calculations, however, the cost of basis construction is significant (since it is not parallelized over elements), we chose smaller buffer sizes when sufficient. In particular, we found buffer sizes of 4.02, 5.10, and 3.24 au (nearest grid points to 4, 5, and 3 au) to be sufficient for CNT, silicene, and fcc Al systems, respectively, and used these in all calculations. In a parallel calculation, this optimization could be omitted if desired by simply using a buffer size of 5 au for all.

\begin{figure}
\subfloat[Energy]{\includegraphics[keepaspectratio=true,width=0.245\textwidth]{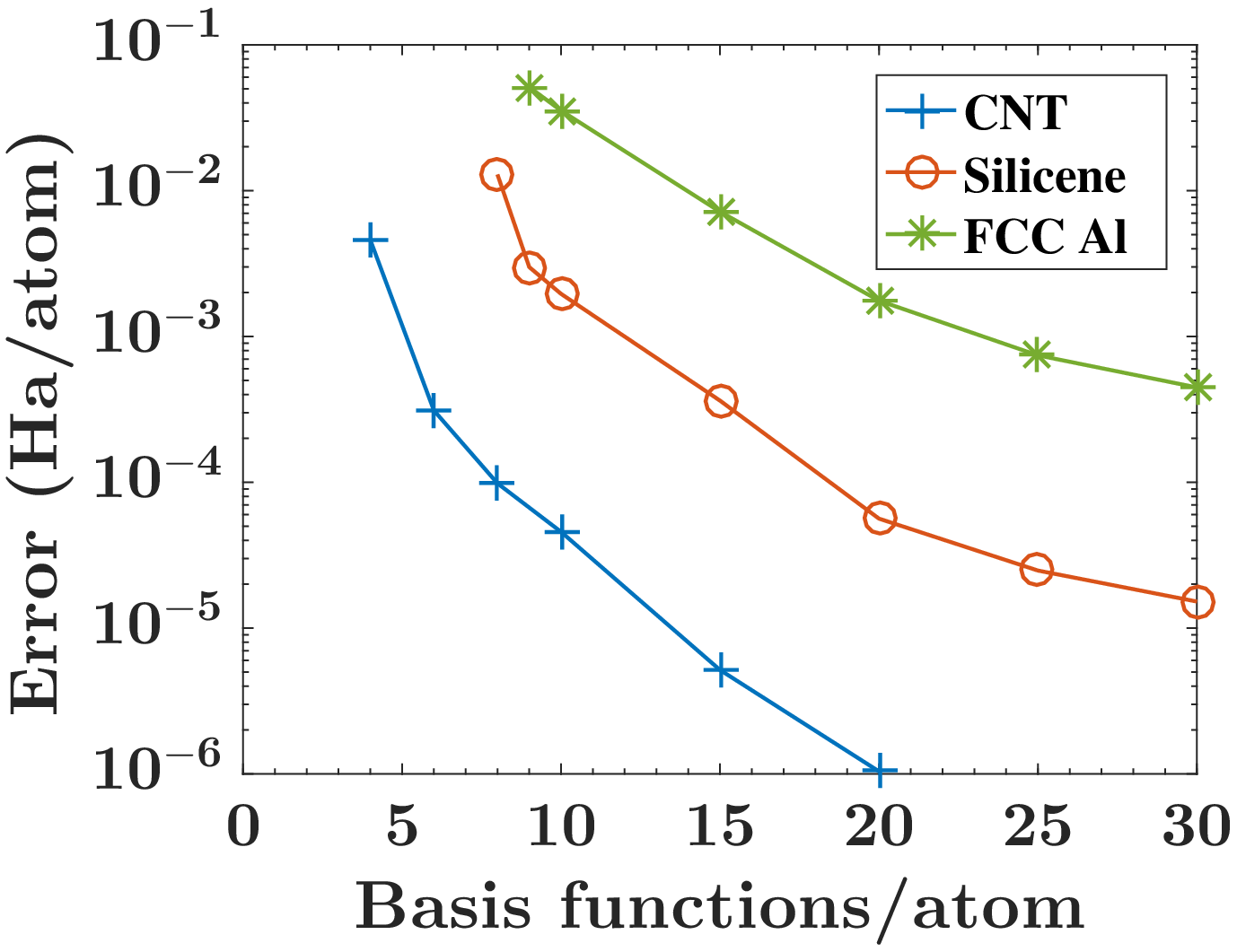}} 
\subfloat[Atomic forces]{\includegraphics[keepaspectratio=true,width=0.245\textwidth]{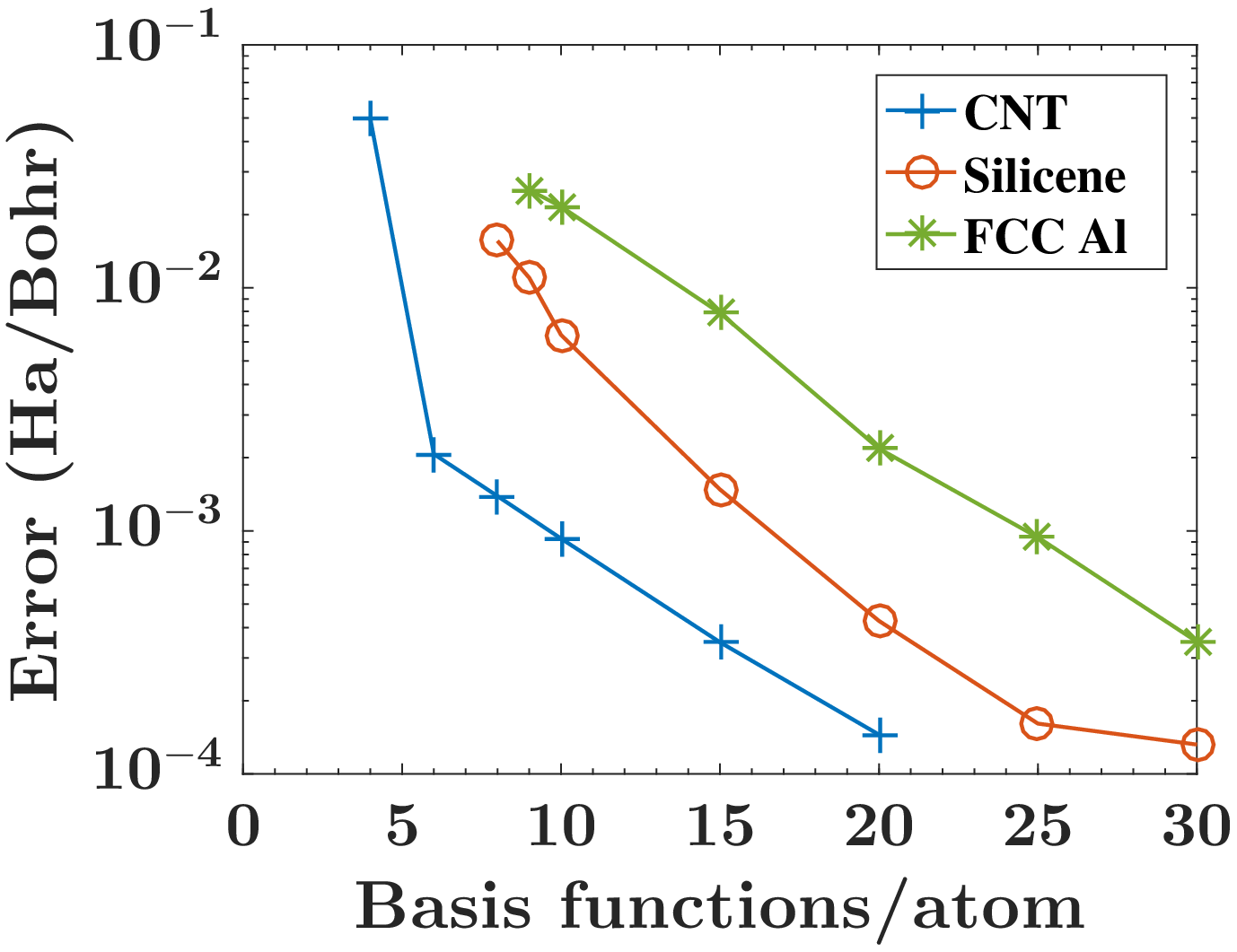}}
\caption{\label{Fig:Convergence:nlb} Convergence of energy and atomic forces with respect to the number of basis functions/atom $\hat{N}_b$ in the DDBP method. The results obtained without projection are used as reference.}
\end{figure}

Figure~\ref{Fig:Convergence:nlb} shows the convergence of DDBP total energies and forces to standard real-space results  
for CNT, silicene, and aluminum systems with $N_a=36$, $36$, and $108$ atoms, respectively. We observe near exponential convergence with respect to $\hat{N}_b$ for all three systems with increasing prefactor as the dimension of the partition $D$ is increased. This is a consequence of the ability to apply exact boundary conditions in more directions for lower-dimensional partitions. Significantly, we see that just $\hat{N}_b \sim 8, 15, \text{ and } 25$ basis functions/atom are sufficient to achieve chemical accuracy in both energy and atomic forces for the CNT, silicene, and aluminum systems, respectively. This is to be compared with $\sim$ 19763, 3488, and 432 grid points/atom, respectively, required by the standard real-space method without projection. Projection thus reduces the dimension of the eigenproblem that must be solved by 1--3 orders of magnitude.  

\begin{table}
\caption{\label{Table:Eig} Representative spectral widths (in Hartree) of full ($\mathcal{H}_d$) and reduced ($\mathcal{H}_p$) Hamiltonians.}
\begin{ruledtabular}
\begin{tabular}{cccc}
                           & CNT         & silicene   & fcc Al        \\  
\hline
$\mathcal{H}_d$    & $265.80$    & $66.32$    & $25.17$               \\
$\mathcal{H}_p$    & $26.22$     & $14.50$    & $10.86$               \\
\end{tabular}
\end{ruledtabular}
\end{table} 

Next, to get a sense of the performance gains to be expected in practice from the substantial reduction in eigenproblem dimension, corresponding reduction in spectral width, and ability to use dense algebra, we examine timings for the CNT, silicene, and aluminum systems. Specifically, we employ $\hat{N}_b=8, 15, \text{ and } 25$ projection basis functions/atom, respectively, to ensure chemical accuracy. When solving the full real-space eigenproblem \eqref{Eq:Eigenproblem:FD}, we employ Chebyshev polynomial degrees of $40$, $20$, and $20$ for the CNT, silicene, and aluminum systems, respectively. For the reduced eigenproblem \eqref{Eq:ProjectedEigenproblem} a degree of $10$ is sufficient in all cases due to the significantly reduced spectral width of the reduced Hamiltonian (Table \ref{Table:Eig}). We note that the reduction in spectral width is most pronounced for finer meshes for which full-Hamiltonian spectral widths are largest. The present projection approach can thus be expected to yield the most significant gains in execution time for systems involving one or more ``hard'' atoms, such as first-row and transition elements.

\begin{figure}
\subfloat[CNT]{\includegraphics[keepaspectratio=true,width=0.245\textwidth]{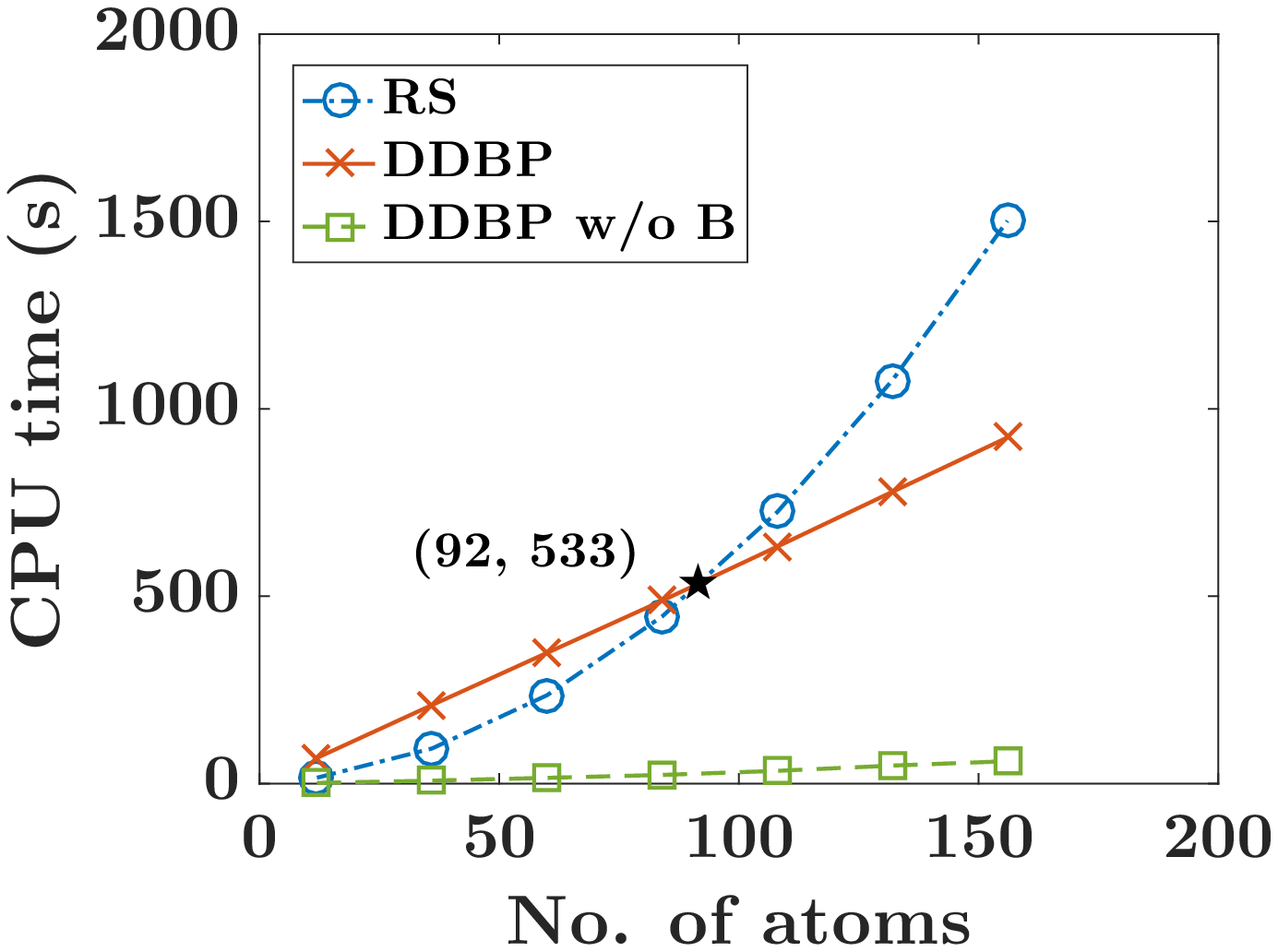}} 
\subfloat[silicene]{\includegraphics[keepaspectratio=true,width=0.245\textwidth]{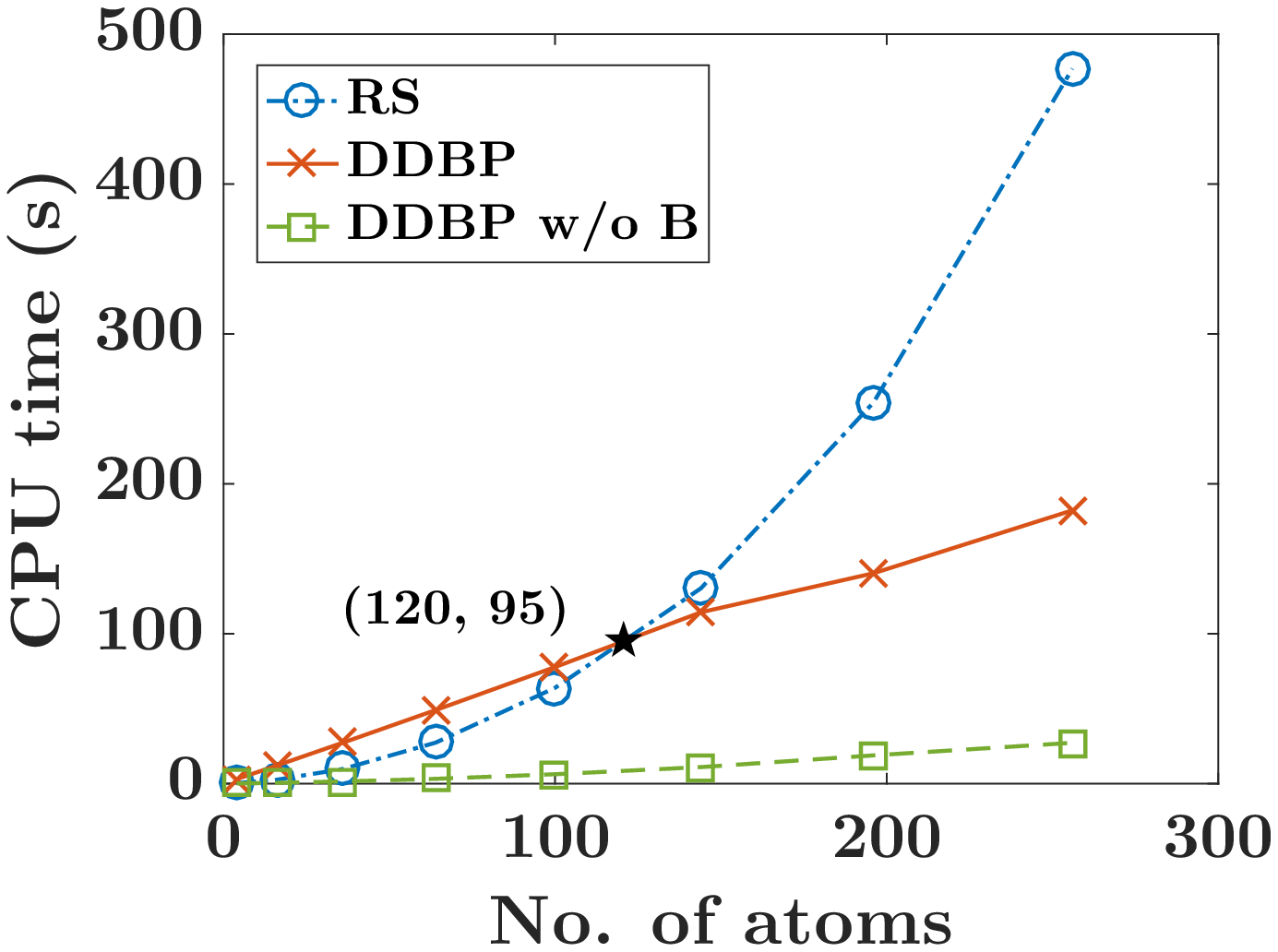}} \\
\subfloat[fcc aluminum]{\includegraphics[keepaspectratio=true,width=0.245\textwidth]{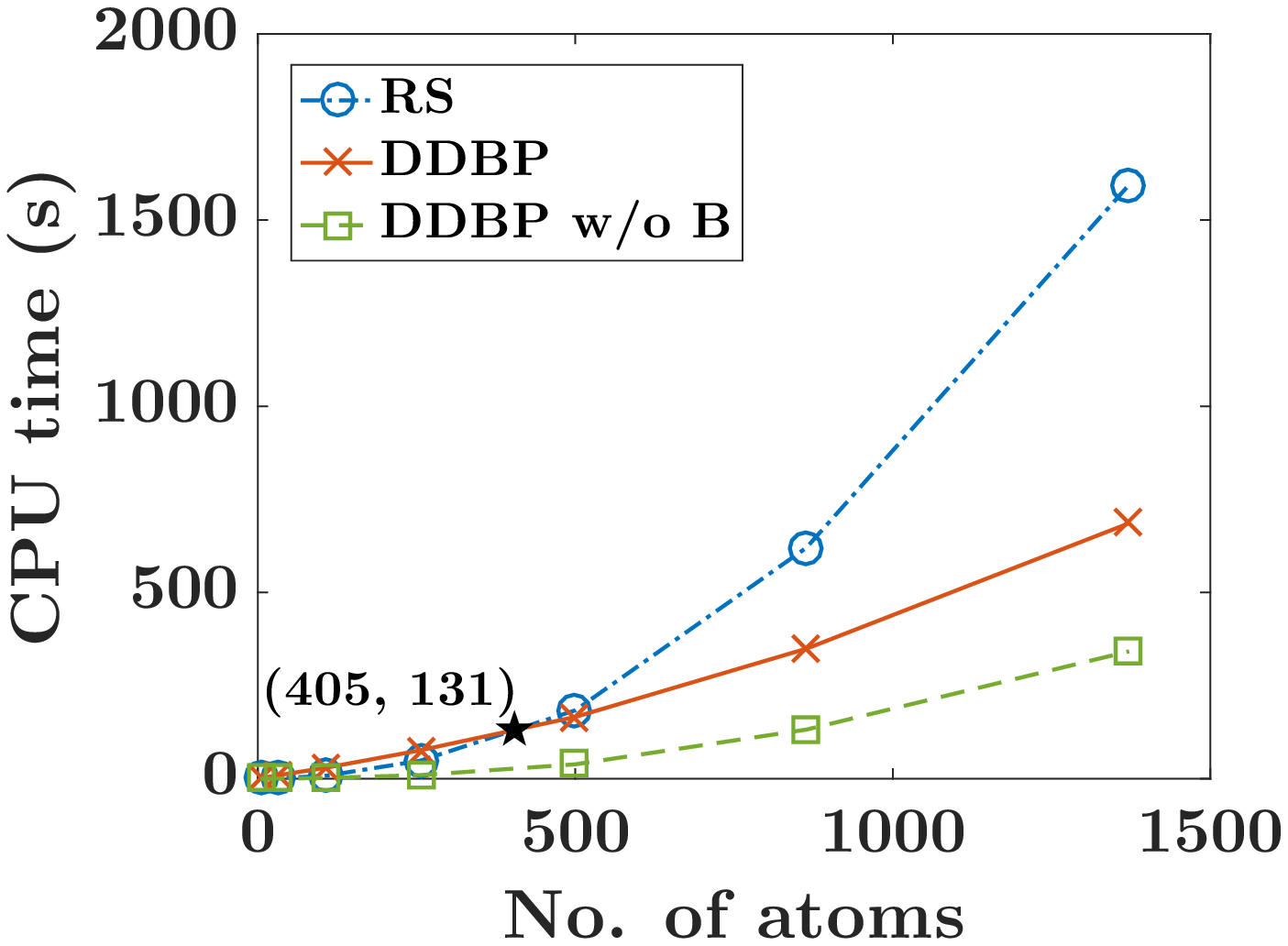}} 
\subfloat{\includegraphics[keepaspectratio=true,width=0.235\textwidth]{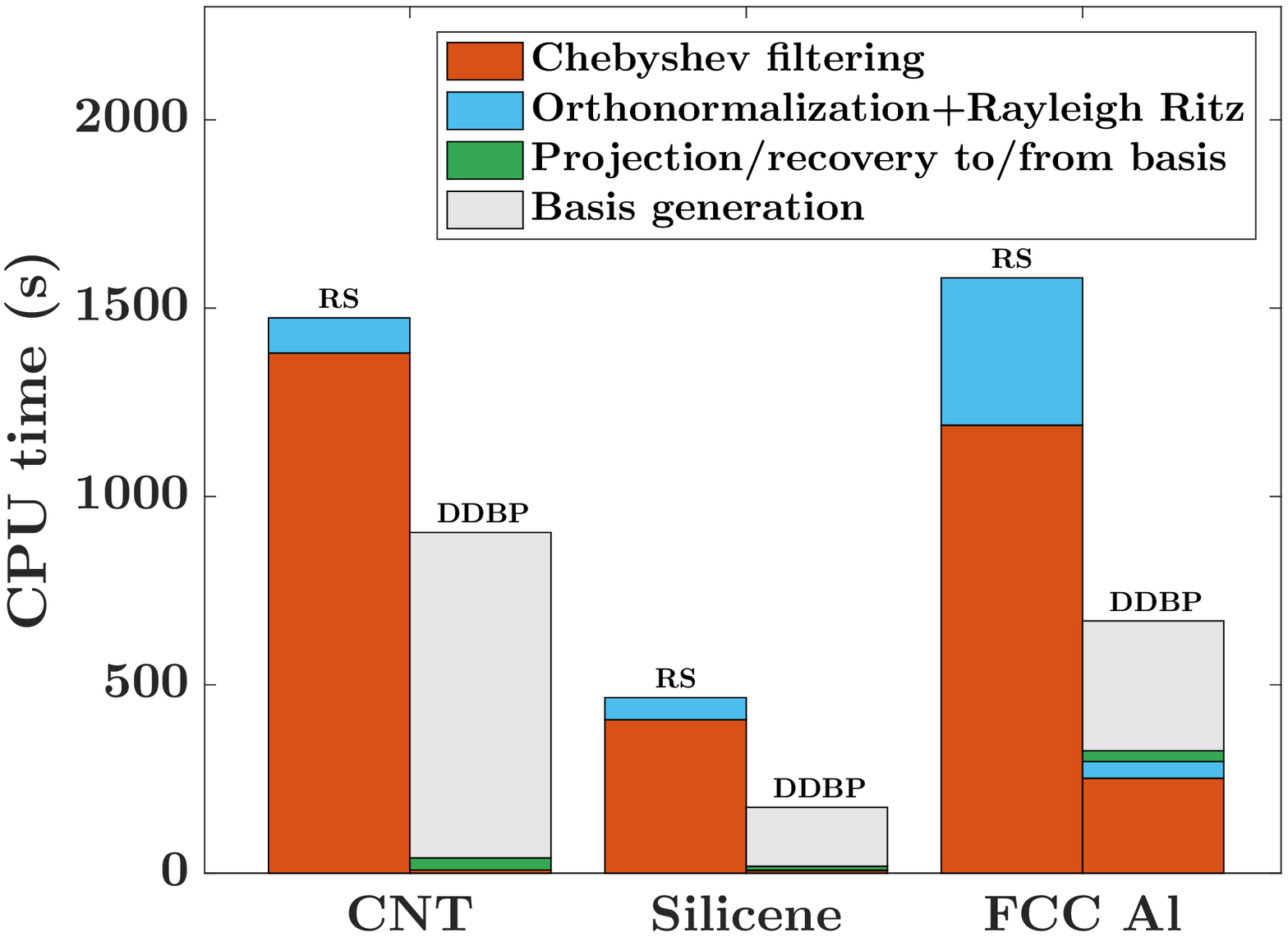}}
\caption{\label{Fig:time:comparison} Time per SCF iteration without (RS) and with (DDBP) projection. \texttt{DDBP w/o B} excludes the time for basis generation, which becomes negligible as problem size and/or number of processors increases. The bar graph shows the timing split for the largest CNT, silicene, and aluminum systems.}
\end{figure}

Figure~\ref{Fig:time:comparison} shows the timings obtained using our serial prototype implementation with and without projection as a function of system size. We see that, even for the serial implementation, projection yields significant speedups as system sizes increase, with factors of 1.6, 2.7, and 2.4 for the largest CNT, silicene, and aluminum systems considered, respectively. However, since basis generation is both $\mathcal{O}(N)$ scaling and naturally parallel, the time for basis generation will vanish as problem size and/or number of processors increases. Hence, to get a better sense for the gains to be expected in a parallel implementation, we show also in Fig.~\ref{Fig:time:comparison} the execution time excluding basis generation, yielding factors of 36.3, 25.0, and 4.9 for the largest CNT, silicene, and aluminum systems considered, respectively. Of particular note is the factor of 4.9 speedup for the bulk Al system since, having a soft potential and correspondingly coarse mesh, speedups for more typical potentials and meshes may be expected to be substantially greater.

Finally, to illustrate the utility of DDBP in challenging physical applications, we compute the vacancy formation energy (VFE) in silicene. Since such calculations must resolve small differences in large energies, they pose a stringent test. For this purpose, we choose a $160$-atom supercell of silicene and create a vacancy, leaving three dangling bonds. Using the same computational parameters in DDBP as in the preceding silicene calculations, we find a VFE of 2.94 eV, within 2~mHa of the full real-space result. While such accuracy is often sufficient, it can be increased further if desired by augmenting the projection basis. For example, increasing from 15 to 25 basis functions/atom, brings agreement to within 0.2~mHa of the full real-space result.


In summary, we have presented an approach to accelerate real-space electronic structure methods several fold, without loss of accuracy, by systematically reducing the dimension of the discrete eigenproblem that must be solved, via projection in a highly efficient discontinuous basis. Upon solving the reduced problem, the full-matrix eigenvalues and vectors are readily recovered and the remainder of the real-space calculation proceeds unmodified. The result is a method which retains the essential simplicity, systematic convergence, and scalability of real-space methods while reducing computational cost substantially. In calculations of quasi-1D, quasi-2D, and bulk metallic systems, we find that accurate energies and forces are obtained with 8--25 projection basis functions per atom, reducing the dimension of full-matrix eigenproblems by 1--3 orders of magnitude.

While the prototype serial implementation employed here already demonstrates significant gains, more significant gains will come with parallel implementation since the basis construction is both linear-scaling and naturally parallel. Basis construction time will then vanish as problem size and/or number of processors increases. This is the focus of current work. On the other hand, with basis construction and filtering costs so reduced, the Rayleigh-Ritz step in the solution of the reduced problem will become apparent at smaller system sizes than in the solution of the full problem wherein it dominates beyond a few thousand atoms \cite{banerjee2016chebyshev}. Avenues to address this include recently developed spectrum slicing \cite{schofield2012specslic} and complementary subspace \cite{banerjee2017cs} approaches which replace the dense Rayleigh-Ritz solution with better-scaling iterative alternatives. Such a compact, sparse reduced Hamiltonian is also well suited to linear-scaling solution methods. Finally, we note that the present reduction strategy might be employed to accelerate other electronic structure methods as well via transforms to and from real space. These offer a number of opportunities for further advances.


This work was performed, in part, under the auspices of the U.S.~Department of Energy by Lawrence Livermore National Laboratory under Contract DE-AC52-07NA27344. Support for this work was provided through the Scientific Discovery through Advanced Computing (SciDAC) program funded by the U.S.~Department of Energy, Office of Science, Advanced Scientific Computing Research and Basic Energy Sciences.




\end{document}